\documentclass[aps,prb,superscriptaddress,floatfix,twocolumn,showpacs]{revtex4-1}
\usepackage{graphicx,graphics}
\usepackage{dcolumn}
\usepackage{amsmath,amssymb,amsfonts}
\usepackage{latexsym,verbatim}
\usepackage{bm}
\usepackage{color}
\usepackage[breaklinks=true,colorlinks,citecolor=blue,linkcolor=blue,urlcolor=blue]{hyperref}

\usepackage{fancybox}
\usepackage{tabularx}

\begin{document}

\author{Alessandro Principi}
\affiliation{Radboud University, institute for Molecules and Materials, NL-6525 AJ Nijmegen, The Netherlands}

\author{Mikhail I. Katsnelson}
\affiliation{Radboud University, institute for Molecules and Materials, NL-6525 AJ Nijmegen, The Netherlands}

\author{Giovanni Vignale}
\affiliation{Department of Physics and Astronomy, University of Missouri, Columbia, Missouri 65211,~USA}

\title{Edge pseudo-magnetoplasmons}

\begin{abstract}
We study the properties of edge plasmons in two-component electron liquids in the presence of pseudomagnetic fields, which have opposite signs for the two different electronic populations and therefore preserve the time-reversal symmetry. The physical realizations of such systems are many. We discuss the cases of strained graphene and of electrons in proximity to a Skyrmion lattice, solving the problem with the Wiener-Hopf technique. We show (i) that two charged counter-propagating acoustic edge modes exist at the boundary and (ii) that, in the limit of large pseudomagnetic fields, each of them involves oscillations of only one of the two electronic components. We suggest that the edge pseudo-magnetoplasmons of graphene can be used to selectively address the electrons of one specific valley, a feature relevant for the emerging field of valleytronics. Conversely, the spin-polarized plasmons at the boundary of Skyrmion lattices can be exploited for spintronics applications. Our solution highlights new features missing in previous (similar) results obtained with uncontrolled approximations, namely a logarithmic divergence of the plasmon velocity, and the absence of gapped edge modes inside the bulk-plasmon gap.
\end{abstract}

\pacs{73.22.Pr,12.39.Dc,73.20.Mf}

\maketitle

\section{Introduction}
Nanoplasmonics,~\cite{Stockman_opt_express_2011} which aims at compressing electromagnetic radiation to sub-wavelength scales by coupling it to matter waves, has recently experienced a strong revival.~\cite{Grigorenko_nature_phot_2012,Brolo_nature_phot_2012,Brongersma_nature_nanotech_2015} Among the reasons, the discovery of two-dimensional (2D) materials has played a major role.~\cite{Novoselov_nature_2005,Novoselov_pnas_2005,Geim_Novoselov_2007,Geim_nature_2013,Kretinin_nano_lett_2014,Caldwell_nature_mater_2015,castroneto_rmp_2009,Das_Sarma_rmp_2011,kotov_rmp_2012,Principi_prb_2014,Woessner_nature_mater_2015} Due to their quasi-2D nature, the atomically-thin layers of van-der-Waals solids exhibit many remarkable and intriguing optical properties:~\cite{castroneto_rmp_2009,Das_Sarma_rmp_2011,kotov_rmp_2012,Mak_nature_phot_2016} they naturally allow to confine the radiation at a surface by coupling it with mobile electrons, thus forming surface-plasmon polaritons.~\cite{Grigorenko_nature_phot_2012,Chen_nature_2012,Fei_nature_2012,Woessner_nature_mater_2015} 
In this respect, graphene has attracted a lot of interest, especially for its record-high plasmon lifetimes:~\cite{Principi_prb_2014,Woessner_nature_mater_2015} plasmon losses have indeed represented so far the fundamental bottleneck for nanoplasmonic applications.~\cite{Khurgin_nature_nanotech_2015}

When a perpendicular magnetic field is applied to a 2D charged liquid, 
collective modes localized at the edge naturally arise.~\cite{Volkov_jetp_lett_1985,Fetter_prb_1985,Giuliani_and_Vignale} These ``edge magnetoplasmons'' have a linear low-energy dispersion, and are decoupled from the (gapped) bulk modes.~\cite{Giuliani_and_Vignale} 
Such modes can be extremely long lived thanks to the strong confinement at the edge and their quasi-one-dimensionality,~\cite{Kumada_prl_2014} 
and  have been extensively studied in the past.~\cite{Volkov_jetp_lett_1985,Fetter_prb_1985} Fetter~\cite{Fetter_prb_1985} calculated their dispersion in a two-dimensional electron gas (2DEG), even though its analytical solution exploited an uncontrolled approximation. Later he solved the problem of edge plasmons in the presence of nearby grounded metal plates by numerical methods.~\cite{Fetter_prb_1986} Notably, edge magnetoplasmons can propagate in both directions along the edge, {\it i.e.} they are not chiral in a strict sense. However, chirality is still present since the ``wrong-direction'' plasmon is gapped, and its gap frequency increases with the magnetic field.~\cite{Fetter_prb_1985}

In many systems, electrons experience {\it pseudo}-magnetic fields, whose main characteristic is to preserve the global time-reversal symmetry. This is the case, e.g., of strained graphene.~\cite{Guinea_nature_phys_2010,Guinea_prb_2010,Vozmediano_phys_rep_2010} Strain, modifying the hopping parameters, enters the low-energy Hamiltonian as a vector potential ${\bm A}({\bm r},t)$. The global time-reversal invariance is assured by the fact that ${\bm A}$ has opposite signs on the two inequivalent valleys (${\bm K}$ and ${\bm K}'$) of the Brillouin zone. In spite of this, Landau quantization has been observed in strained samples and the effective magnetic field has been shown to reach values of hundreds of Tesla.~\cite{Levy_science_2010} This system is, from the point of view of the energy spectrum, equivalent to a quantum Hall insulator. 

\begin{figure}[t]
\begin{center}
\begin{tabularx}{\columnwidth}{X X}
\noindent\parbox[c]{\hsize}{\includegraphics[width=0.49\columnwidth]{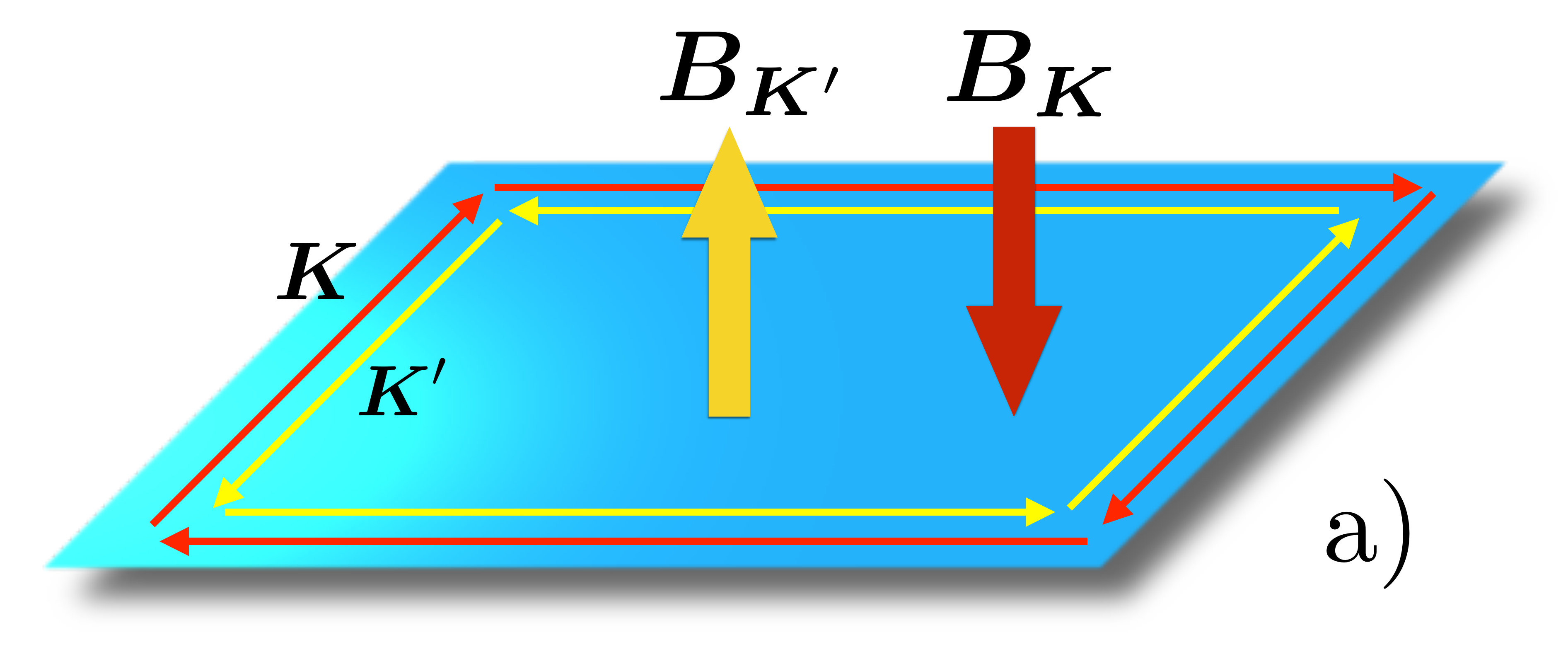}}
&
\noindent\parbox[c]{\hsize}{\includegraphics[width=0.49\columnwidth]{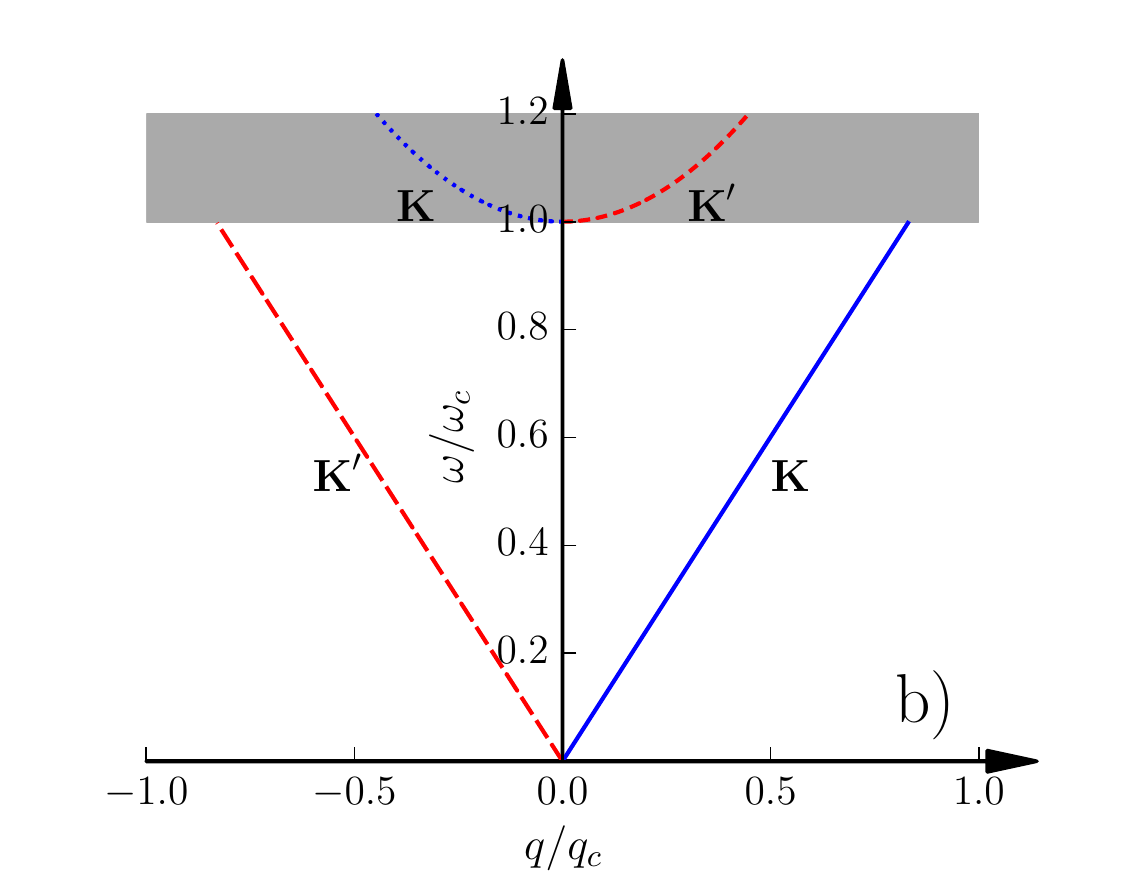}}
\end{tabularx}
\end{center}
\caption{
Panel a) a schematic view of the theoretical model: the two electronic components experience opposite pseudomagnetic fields. In the case of graphene this is achieved by strain, which acts as an effective magnetic field with opposite signs for the electrons in the two valleys (${\bm K}$ and ${\bm K}'$). Two counter-propagating plasmons appear at the edge of the system, each of them mainly due to density oscillations in a specific valley.
Panel b) the dispersion of edge collective modes in units of the cyclotron frequency $\omega_c$, as a function of the momentum $q$ measured in units of $q_c = (k_{\rm F}\ell^2)^{-1}$ [$\ell=\sqrt{c/(eB)}$ is the magnetic length]. We set the filling factor $\nu=1$ ($v_{\rm p}\simeq 1.2 v_{\rm F}$). Each electronic component, depending on the range of frequencies explored, can support up to two charged collective modes, one of which lives inside the gap of the particle-hole continuum (shaded region). The second mode is always gapped, with zero-momentum energy $\hbar\omega_c$.
\label{fig:one}}
\end{figure}

Naively, when doping is sufficiently high and inter-valley scattering is neglected, one would expect the electrons of each valley to behave as a 2DEG subject to an effective magnetic field with a well defined direction [Fig.~\ref{fig:one}a)].
Therefore, each valley should exhibit {\it two} edge plasmons, one of which gapped [Fig.~\ref{fig:one}b)], with the direction of propagation of the acoustic plasmon determined by the sign of the pseudomagnetic field in the given valley. Therefore, at low frequency one expects two counter-propagating edge magnetoplasmons to emerge, each due to density oscillations of one of the two electronic components. Unfortunately, the problem is not so simple: even neglecting direct scattering between them, the two valleys are always electrostatically coupled, and a density fluctuation in one of them will invariably influence the electrons in the other. This fact makes the problem completely non-trivial and, since one of the two valleys is always off-resonance ({\it i.e.} it experiences an effective magnetic field with the wrong sign), it could in principle destroy the collective modes. We find that the two counter-propagating acoustic edge plasmons survive, but that the valleys are not completely disentangled. Each collective mode stems indeed from the superposition of density oscillations in {\it both} valleys, and becomes ``localized'' in one of them only in the limit of large pseudomagnetic fields. We stress again that high field values are actually attainable in experiments.

A similar situation occurs when electrons travel in proximity to a Skyrmion crystal realized, e.g., in a chiral magnet. The complex, topological magnetic structure is responsible for the emergence of an ``effective electrodynamics'' determined by the Berry phases accumulated by the electrons moving in it.~\cite{Schulz_nature_phys_2012} 
Traveling in such a structure the electron spin tends to adiabatically align with the direction of the magnetization. At suitable electronic densities, spin-up electrons accumulate a Berry phase which has a sign opposite to that of spin-down electrons.~\cite{Bruno_prl_2004} As a consequence, the two spin populations ``skew'' in opposite directions.
In this case the pseudomagnetic field ($\sim 2.5~{\rm T}$ in ${\rm MnSi}$~\cite{Schulz_nature_phys_2012}) originates from the Berry phase accumulated by electrons passing through the complex magnetic structure: this system offers therefore a beautiful model to study the impact of the (real-space) Berry phase on collective modes.

In this letter we solve the edge-plasmon problem in a two-component 2D electronic system subject to a pseudomagnetic field. 
We solve the full Wiener-Hopf problem~\cite{Noble_WH_book} defined by constitutive equations and electrostatics, and we provide a comparison with an approximate solution {\it \`a la} Fetter. 

\begin{figure}[t]
\begin{center}
\begin{tabularx}{\columnwidth}{X X}
\multicolumn{2}{c}{ \includegraphics[width=0.99\columnwidth]{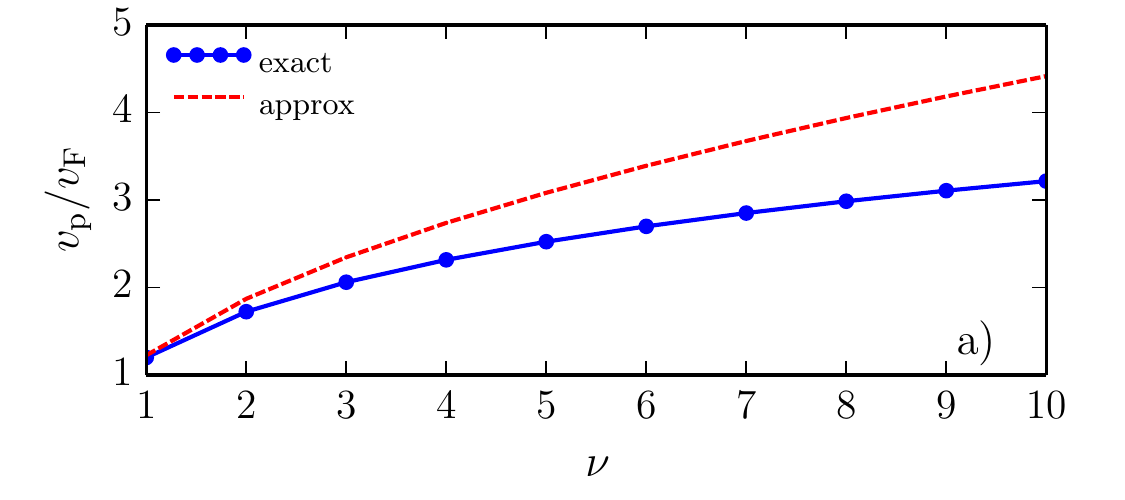} }
\\
\multicolumn{2}{c}{ \includegraphics[width=0.99\columnwidth]{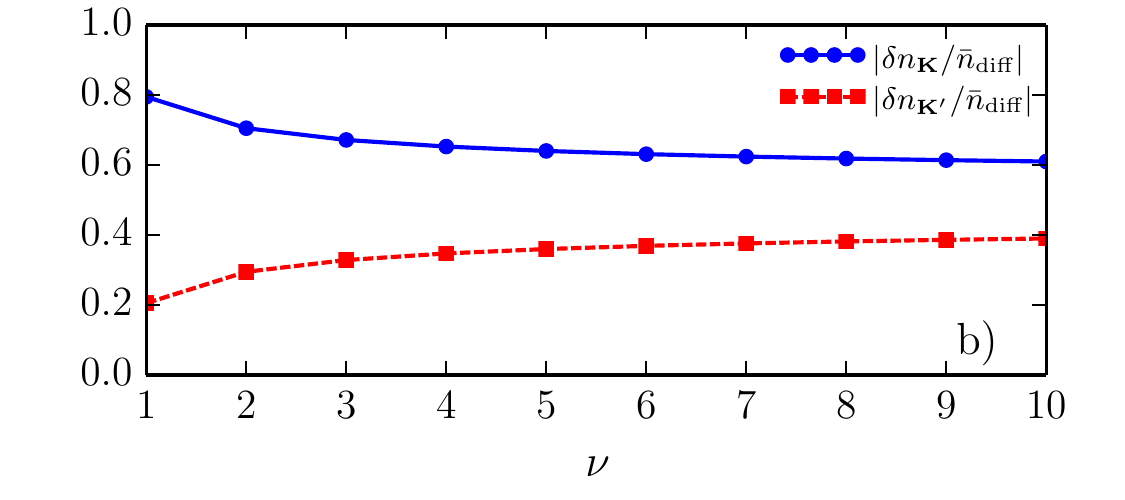} }
\end{tabularx}
\end{center}
\caption{
Panel a) the sound velocity of the acoustic edge pseudo-magnetoplasmon $v_{\rm p} = \omega_{\rm p}(q)/q$ in units of the Fermi velocity, plotted as a function of the filling factor $\nu$. The dots represent the Wiener-Hopf result, while the dashed line is the solution approximated {\it \'a-la} Fetter [see Eq.~(\ref{eq:sound_vel_approx})]. We cut-off the logarithmic divergence of $v_{\rm p}$ by setting ${\bar q} = 0.01$.
Panel b) the degree of valley polarization of the right-moving edge pseudo-magnetoplasmon, given by $|\delta n_{\bm K}/{\bar n}_{\rm diff}| = (v_{\rm p}+ s)/(2v_{\rm p})$ and $|\delta n_{{\bm K}'}/{\bar n}_{\rm diff}| = (v_{\rm p} - s)/(2 v_{\rm p})$. Note that at large magnetic field ($\nu=1$), $80\%$ of the contribution to density oscillations comes from electrons in valley ${\bm K}$, and only $20\%$ from those living around the ${\bm K}'$ point. For the left-moving edge plasmon an analogous figure can be drawn with valleys ${\bm K}$ and ${\bm K}'$ interchanged.
\label{fig:two}}
\end{figure}

\section{The model}
For the sake of definiteness we consider a strained graphene sheet which occupies the half plane $x<0, z=0$
We assume that the presence of the edge does not affect the low-energy physics of the system, and that the electrons can be described by the massless Dirac fermion Hamiltonian~\cite{castroneto_rmp_2009,Das_Sarma_rmp_2011,kotov_rmp_2012}
\begin{eqnarray}
{\cal H}_0 = v_{\rm F}\sum_{{\bm k},\alpha,\beta} {\hat \psi}^\dagger_{{\bm k},\alpha} ({\bm k}+ {\bm A})\cdot{\bm \sigma}_{\alpha\beta} {\hat \psi}_{{\bm k},\beta}
~,
\end{eqnarray}
where ${\hat \psi}^\dagger_{{\bm k},\alpha}$ (${\hat \psi}_{{\bm k},\alpha}$) creates (destroys) a particle with momentum ${\bm k}$ and pseudospin $\alpha$, $v_{\rm F}$ is the Fermi velocity, 
$A_x =  \xi \beta (u_{xx}-u_{yy})/a$ 
and
$A_y = - 2\xi \beta u_{xy}/a$
are the two component of the pseudomagnetic vector potential generated by the strain tensor ${\bm u}_{ij}({\bm r})$ [here $\beta = -\partial \ln(t)/\partial \ln(a) \simeq 2$, $a = 1.4~{\rm \AA}$, $\xi$ is a numerical constant of order one].~\cite{Guinea_nature_phys_2010,Guinea_prb_2010,Vozmediano_phys_rep_2010} In what follows we assume that the shape of the applied strain is such that ${\bm \nabla} \times {\bm A} = \pm B {\hat {\bm z}}$, where the pseudomagnetic field $B$ is constant, while the plus (minus) sign applies to electrons in valley ${\bm K}$ (${\bm K}'$). Even though the strain field must have a trigonal symmetry to induce a constant pseudomagnetic field,~\cite{Guinea_nature_phys_2010,Guinea_prb_2010,Vozmediano_phys_rep_2010} we assume that the curvature of the edge is negligible, and we treat it as a straight line. We neglect inter-valley scattering, assume graphene to be in the Fermi-liquid regime,~\cite{Giuliani_and_Vignale} and we study the electronic transport by means of linearized hydrodynamic equations.~\cite{Principi_prb_2016,Bandurin_science_2016,Crossno_science_2016} The electron densities in each valley are separately conserved, and satisfy the continuity equations
\begin{eqnarray} \label{eq:continuity}
&& \partial_t \delta n_{{\bm K}} + n_0 {\bm \nabla}\cdot {\bm v}_{{\bm K}} = 0
~,
\nonumber\\
&& \partial_t \delta n_{{\bm K}'} + n_0 {\bm \nabla}\cdot {\bm v}_{{\bm K}'} = 0
~,
\end{eqnarray}
where $\delta n_{{\bm K}}$ ($\delta n_{{\bm K}'}$) is the non-equilibrium density fluctuation in valley ${\bm K}$ (${\bm K}'$), while $n_0$ is its equilibrium value. Hereafter we suppress space and time indices for brevity. The electron velocities ${\bm v}_{{\bm K}}$ and ${\bm v}_{{\bm K}'}$ obey the Navier-Stokes equations~\cite{Landau_6,Giuliani_and_Vignale}
\begin{eqnarray} \label{eq:Navier_Stokes}
&& \partial_t {\bm v}_{{\bm K}} + \omega_c {\hat {\bm z}} \times {\bm v}_{{\bm K}} + \frac{s^2}{n_0} {\bm \nabla}\delta n_{{\bm K}} - \frac{e}{m} {\bm \nabla} \phi = 0
~,
\nonumber\\
&& \partial_t {\bm v}_{{\bm K}'} - \omega_c {\hat {\bm z}} \times {\bm v}_{{\bm K}'} + \frac{s^2}{n_0} {\bm \nabla}\delta n_{{\bm K}'} - \frac{e}{m} {\bm \nabla} \phi = 0
~,
\end{eqnarray}
where $m = \hbar k_{\rm F}/v_{\rm F}$ is the cyclotron mass ($k_{\rm F}$ is the Fermi momentum), $\omega_{\rm c} = e B/(m c)$ is the classical cyclotron frequency, and $s= \sqrt{m^{-1} \partial P/\partial n} = v_{\rm F}/\sqrt{2}$.~\cite{Giuliani_and_Vignale} 
Finally, the electrostatic potential is given by
\begin{eqnarray} \label{eq:potential}
\phi ({\bm r}) = e \int d^2{\bm r}' \frac{\delta n_{{\bm K}}({\bm r}') + \delta n_{{\bm K}'} ({\bm r}')}{|{\bm r}-{\bm r}'|}
~.
\end{eqnarray}
Since the translational invariance along the ${\hat {\bm y}}$ direction is not broken, all functions have a dependence of the form $e^{-i(\omega t - q y)}$. 
Eqs.~(\ref{eq:continuity})-(\ref{eq:potential}) constitute a system of integro-differential equations that can be solved using the Wiener-Hopf technique.~\cite{Noble_WH_book} We calculate the sound velocity of the two counter-propagating edge pseudo-magnetoplasmons. Furthermore, we show that in the limit $B\to \infty$ the two valleys decouple and each collective mode is due to density oscillations of only one of them. 

Fetter~\cite{Fetter_prb_1985} simplified the problem by introducing an  approximation of Eq.~(\ref{eq:potential}), replacing it with
\begin{eqnarray} \label{eq:potential_diff}
\partial_x^2 \phi(x) - 2 q^2 \phi(x) = 4\pi e |q| \big[ \delta n_{{\bm K}}(x) + \delta n_{{\bm K}'} (x) \big]
~.
\end{eqnarray}
The big advantage of Eq.~(\ref{eq:potential_diff}) is that, while leaving intact the first two moments of the interaction potential integrated across the edge, it allows to study a system of ordinary linear differential equations. However, effects that depend on the long range of the interaction {\it along} the edge are in this way lost. Note indeed that the asymptotic behavior of Eq.~(\ref{eq:potential_diff}) in the limit $q\to 0$ is completely different from that of the Fourier transform of Eq.~(\ref{eq:potential}). Below, we compare our exact results with those obtained with the approximation~(\ref{eq:potential_diff}).  We stress that the solution obtained with the Wiener-Hopf method in not just an incremental improvement of Fetter's result, but reveals features missing in the approximate result. Namely, (i) the logarithmic divergence of the plasmon velocity at small momenta due to the long-range nature of the Coulomb interaction,~\cite{Volkov_jetp_lett_1985} and (ii) the absence of gapped modes with energy below $\hbar \omega_c$.
The details of the calculation in the approximate model, which closely parallels Fetter's derivation,~\cite{Fetter_prb_1985} are given in the Supplemental Online Material.

\section{The Wiener-Hopf solution}
To solve the problem posed by Eqs.~(\ref{eq:continuity})-(\ref{eq:potential}), we first introduce $n_{{\rm sum}({\rm diff})}(x) \equiv \delta n_{{\bm K}}(x) \pm \delta n_{{\bm K}'}(x)$. The resulting equation for $n_{\rm diff}(x)$ is independent of $\phi(x)$, and its solution reads $n_{\rm diff}(x) = {\bar n}_{\rm diff} e^{\kappa_- x}$, where $\kappa_- = \sqrt{q^2 + s^{-2}(\omega_c^2 - \omega^2)}$. ${\bar n}_{\rm diff}$ is a constant to be determined from the boundary conditions.
Plugging this solution back into Eqs.~(\ref{eq:continuity})-(\ref{eq:potential}), and taking their one-sided Fourier transform,~\cite{Noble_WH_book} we find
\begin{subequations} \label{eq:n_plus_k_final_main}
\begin{eqnarray}
\label{eq:n_plus_k_final_main_a}
&& n_{\rm sum}(k) =  \frac{\displaystyle \frac{2e n_0}{m s^2} (k^2 + q^2) \phi(k) 
+
\left[ i k  +\frac{q^2}{\kappa_-} \frac{\omega_c^2}{\omega^2} \right] 
{\bar \Xi}}{k^2 + \kappa_-^2}
~,
\nonumber\\
\\
\label{eq:n_plus_k_final_main_b}
&& {\bar n}_{\rm diff} = \frac{q}{\kappa_-} \frac{\omega_c}{\omega} {\bar \Xi}
~.
\end{eqnarray} 
\end{subequations}
Here we used the fact that $v_{{\bm K},x}(0)=v_{{\bm K}',x}(0) = 0$, and we defined ${\bar \Xi} \equiv n_{\rm sum}(0)  - 2 e n_0 \phi(0)/(m s^2)$.
We extend $k$ to the whole complex plane, and we denote with the subscript ``$+$'' [``$-$''] functions that are analytic in the upper [lower] half. The functions $n_{\rm sum}(k)$ and $\phi(k)$ in Eq.~(\ref{eq:n_plus_k_final_main_a}) are, by construction, analytic for $\Im m(k)\geq 0$,~\cite{Noble_WH_book} and we therefore rename $\phi(k)\to \phi_+(k)$ and $n_{\rm sum}(k) \to n_+(k)$.
Analyticity requires that the numerator of Eq.~(\ref{eq:n_plus_k_final_main_a}) vanishes for $k=i\kappa_-$: we will make use of this condition below to determine the plasmon dispersion.
Taking the double-sided Fourier transform of Eq.~(\ref{eq:potential}), noting that the left-hand side is $\phi_+(k)+\phi_-(k)$, and combining it with Eq.~(\ref{eq:n_plus_k_final_main_a})
we get
\begin{eqnarray} \label{eq:phi_equation_main}
(k^2 + \kappa_-^2)G(k)\phi_+(k)+ (k^2 + \kappa_-^2)\phi_-(k) = 2\pi e {\bar \Xi} F(k)
~.
\nonumber\\
\end{eqnarray}
where $G(k) \equiv 1 + 2 \alpha \sqrt{k^2 + q^2}/(k^2 + \kappa_-^2)$, $F(k) = - \big[ i k 
+ q^2 \omega_c^2/(\kappa_- \omega^2)\big]/\sqrt{k^2+q^2}$, and $\alpha = 2 \pi e^2 n_0/(m s^2)$.
Eq.~(\ref{eq:phi_equation_main}) can be solved with the Wiener-Hopf technique. Using a well-known theorem of complex analysis,~\cite{Noble_WH_book} we rewrite $G(k)=G_+(k)/G_-(k)$, where ($\eta \to 0^+$)
\begin{eqnarray} \label{eq:G_pm_main}
G_{\pm}(k) = \exp \left[ \int_{-\infty\mp i\eta}^{\infty\mp i\eta} \frac{dz}{2\pi i} \frac{\ln G(z)}{z-k} \right]
~.
\end{eqnarray}
The function $G_+(k)$ [$G_-(k)$] is analytic in the upper [lower] half of the complex plane. Eq.~(\ref{eq:phi_equation_main}) then becomes
\begin{equation} \label{eq:phi_equation_2_main}
G_+(k) \phi_+(k)+ G_-(k) \phi_-(k) =  2\pi e {\bar \Xi} F(k) \frac{G_-(k)}{k^2 + \kappa_-^2}
~.
\end{equation}
The term on the right-hand side of Eq.~(\ref{eq:phi_equation_2_main}) can be rewritten as
$F(k) G_-(k)/(k^2 + \kappa_-^2) = F_+(k) + F_-(k)$,~\cite{Noble_WH_book}
where $F_+(k)$ [$F_-(k)$] is analytic in the upper [lower] half of the complex plane, and reads
\begin{eqnarray} \label{eq:F_pm_main}
F_\pm (k) = \pm \int_{-\infty \mp i\eta}^{\infty \mp i \eta} \frac{dz}{2\pi i} \frac{F(z)}{z-k} \frac{G_-(z)}{z^2 + \kappa_-^2}
~.
\end{eqnarray}
Eq.~(\ref{eq:phi_equation_2_main}) now reads
\begin{equation} \label{eq:phi_equation_3_main}
G_+(k) \phi_+(k) - 2\pi e F_+(k) {\bar \Xi} = 
2\pi eF_-(k) {\bar \Xi} - G_-(k) \phi_-(k)
~.
\end{equation}
Since the left-hand side is analytic for $\Im m (k) \geq 0$ and the right-hand side is analytic for $\Im m(k)\leq 0$, together they define a function analytic in the whole complex plane. Moreover, both sides of Eq.~(\ref{eq:phi_equation_3_main}) vanish in the limit $|k|\to \infty$. As a consequence,~\cite{Noble_WH_book} they must be separately equal to zero. Therefore,
$\phi_+(k) = 2\pi e {\bar \Xi} F_+(k)/G_+(k)$.
Plugging this back in Eq.~(\ref{eq:n_plus_k_final_main_a}) we finally get
\begin{equation} \label{eq:n_plus_k_final_2_main}
n_{+}(k) =  \left[\displaystyle 2 \alpha (k^2 + q^2) \frac{F_+(k)}{G_+(k)} +  i k 
+\frac{q^2}{\kappa_-} \frac{\omega_c^2}{\omega^2} \right] \frac{{\bar \Xi}}{k^2+\kappa_-^2}
~.
\end{equation} 
Since $n_+(k)$ is, by definition, analytic for $\Im m(k)>0$, the square brackets in Eq.~(\ref{eq:n_plus_k_final_2_main}) has to vanish for $k = i \kappa_-$ in order to cancel the pole in the denominator. 
Performing the integrals~(\ref{eq:G_pm_main}) and~(\ref{eq:F_pm_main}), setting $\omega_{p}(q) = v_{\rm p} q$, and taking the limit $q\to 0$, from the square brackets in Eq.~(\ref{eq:n_plus_k_final_2_main}) we get $s^2/v_{\rm p}^2  - 1 = 2 {\bar \alpha} f/g$, 
where ${\bar \alpha} \equiv s \alpha/\omega_c$, and
\begin{eqnarray}
g &=&
\exp \Bigg[ \frac{2 {\bar \alpha}}{\pi} \int_0^\infty d x 
\frac{x^2 + 1}{(x^2-1)^2 + 4{\bar \alpha}^2 x^2} 
\ln\left(\frac{1+x}{2}\right)
\Bigg]
~,
\nonumber\\
f &=&
\frac{1}{\pi} {\cal P} \int_{{\bar q}}^\infty \frac{d y}{y + 1}\left(\frac{s^2}{v_{\rm p}^2} + y\right) \frac{y^{-1}}{y^2 - 1}
\nonumber\\
&\times&
\exp \Bigg[
-\frac{2 {\bar \alpha}}{\pi} \int_0^\infty d x 
\frac{ \displaystyle (x^2 + 1) \ln\left(\frac{y+x}{y+1}\right) }{(x^2-1)^2 + 4{\bar \alpha}^2 x^2} 
\Bigg]
~.
\end{eqnarray}
Note that ${\bar \alpha} = \sqrt{2} N_{\rm f} \alpha_{\rm ee} (\nu-1/2)$, where $N_{\rm f}$ is the number of residual fermion flavors, $\alpha_{\rm ee} = e^2/(\hbar v_{\rm F})$ the dimensionless coupling constant, and $\nu$ the filling factor (number of filled Landau levels). In the presence of unscreened electron-electron interactions the integral on the second line is infrared divergent in the limit ${\bar q}\to 0$. The edge pseudo-magnetoplasmon velocity therefore diverges as
$v_{\rm p}^2 \to - 2 {\bar \alpha} \ln(q)/(\pi g)$.~\cite{Volkov_jetp_lett_1985}
In the Supplemental Online Material we solve the problem {\it \'a-la} Fetter, by replacing Eq.~(\ref{eq:potential}) with Eq.~(\ref{eq:potential_diff}), and we find
\begin{eqnarray} \label{eq:sound_vel_approx}
v_{\rm p, approx} = s \sqrt{1 + 2\sqrt{2} \frac{2 \pi e^2 n_0}{m s \omega_c}}
~.
\end{eqnarray}
A comparison between $v_{\rm p}$ and the approximate result of Eq.~(\ref{eq:sound_vel_approx}) is given in Fig.~\ref{fig:two}a). In this plot we set ${\bar q} = 0.01$. 

Since the problem is symmetric for $q\to -q$, at any given frequency it is possible to excite two counter-propagating plasmons. Let us now discuss the degree of valley polarization of such plasmons. It is easy to show that the two electronic components oscillate with opposite phases. Therefore $|n_{\rm diff}| > |n_{\rm sum}|$, and the right quantities that display the degree of valley polarization are $|\delta n_{\bm K}/{\rm n}_{\rm diff}|$ and $|\delta n_{{\bm K}'}/{\rm n}_{\rm diff}|$. 
For weak pseudomagnetic fields the two are identical, {\it i.e.} both valleys are involved in plasmon oscillations. However, in the limit $B\to\infty$ one of the two valleys is completely ``frozen'' and the oscillations involve only the other one. In this case, for example, $|\delta n_{\bm K}/{\rm n}_{\rm diff}| \simeq 1$ and $|\delta n_{{\bm K}'}/{\rm n}_{\rm diff}| \simeq 0$.
We derive an explicit expression for these two quantities. Let us first note that when $\omega = \omega_{\rm p}(q)$, ${\bar n}_{\rm diff} = s{\bar \Xi}/v_{\rm p}$ [see Eq.~(\ref{eq:n_plus_k_final_main_b})]. The value of $n_{\rm sum}(x)$ at the boundary is instead obtained by taking the Fourier transform of Eq.~(\ref{eq:n_plus_k_final_2_main}) in the limit $x\to 0^-$. This is shown in detail in the Supplemental Online Material. The resulting expressions are fairly simple, although their numerical integration turns out to be quite challenging. At the same time, the approximate model gives $n_{\rm sum}(0) =s^2 {\bar \Xi}/v_{\rm p}^2$ which is in very good agreement with the result obtained with the Wiener-Hopf technique (when $v_{\rm p}$ is calculated with this method). Using the approximate expression, $|\delta n_{\bm K}/{\rm n}_{\rm diff}| = (v_{\rm p} + s)/(2v_{\rm p})$ and $|\delta n_{{\bm K}'}/{\bar n}_{\rm diff}| = (v_{\rm p} - s)/(2v_{\rm p})$. In Fig.~\ref{fig:two}b) we plot the two functions for the right-propagating mode. Note that at $\nu=1$ the $80\%$ of the contribution to this plasmon comes from the electrons of valley ${\bm K}$, while only the $20\%$ is due to those around the ${\bm K}'$ point.

Finally, it is possible to show that no other mode lives inside the bulk-plasmon gap. Such a mode should have a zero-momentum frequency smaller than $\omega_c$. The plasmon equation for gapped modes is obtained as before by considering the term in the square brackets in Eq.~(\ref{eq:n_plus_k_final_2_main}) and setting $k=i\kappa_-$ and $\omega = \omega_c \Delta$, with $0<\Delta<1$. It is easy to show that in the limit $q\to 0$ the resulting equations are identical to those obtained for the acoustic modes, when these are evaluated in the limit $v_{\rm p} \to \infty$ and ${\bar \alpha} = s \alpha/(\omega_c \sqrt{1 - \Delta^2})$. The plasmon equation is therefore a function of only ${\bar \alpha}$, and it has no solution unless ${\bar \alpha} \to \infty$ (which corresponds to $\Delta=1$). Therefore the gapped mode has a minimum energy equal to $\hbar \omega_c$.

\section{Conclusions}
In this letter we have discussed the problem of collective modes confined at the boundaries of a two-component 2D system subject to a pseudomagnetic field which preserves the time-reversal symmetry.~\cite{Guinea_nature_phys_2010,Guinea_prb_2010,Vozmediano_phys_rep_2010,Schulz_nature_phys_2012,Bruno_prl_2004} 
 This property is ensured by the fact that it has opposite signs for the two different electron populations.
We have shown that (i) two counter-propagating acoustic plasmons live at the edge of the system, and that (ii) in the limit of large pseudomagnetic field the excited plasmon involve only density oscillations of one of the two electronic components. The other one is completely ``frozen''.
We stress that the very same phenomenology emerges in many different physical systems, and therefore the solution we provide has a wide applicability.

In graphene, the edge modes induced by shear strain deformations~\cite{Guinea_nature_phys_2010,Guinea_prb_2010,Vozmediano_phys_rep_2010} are valley-polarized. They can therefore be used to selectively excite electrons in one of the two valleys by, e.g., optical means, by carefully choosing the energy and wavevector of the imparted external perturbation. This fact, similar to the valley-selective circular dichroism of transition metal dichalcogenides.~\cite{Zeng_nature_nanotech_2012}, has a direct impact on the emerging field of valleytronics.~\cite{Rycerz_nature_phys_2007,Zeng_nature_nanotech_2012,Kim_science_2014} 
Furthermore, since the edge plasmon velocity depends on the strain field, it is possible to draw an analogy with the propagation of light in media with different refractive indexes, and imagine to induce focusing, anti-focusing and interference between collective modes by means of a properly chosen strain pattern.~\cite{Guinea_nature_phys_2010,Guinea_prb_2010,Vozmediano_phys_rep_2010,Levy_science_2010}

Our edge pseudo-magnetoplasmons are conceptually different from the edge ``Berry plasmons" recently introduced in Ref.~\onlinecite{Song_PNAS_2016}. The latter are driven by a pseudo-magnetic field in momentum space (Berry curvature) whereas our valley-selective pseudo-magnetic field acts in real space, therefore opening a gap in the spectrum of the modes propagating in the ``wrong'' direction. No such gap is present in the spectrum of Berry plasmons.

Another more ``exotic'' example 
is given by electrons traveling in a Skyrmion lattice.~\cite{Schulz_nature_phys_2012,Bruno_prl_2004} 
Our theory predicts the existence of counter-propagating spin-polarized acoustic plasmons, which can be exploited for spintronics applications.~\cite{Sinova_prl_2004,Mishchenko_prl_2004,Inoue_prbr_2004}

\section{Acknowledgments}
A.P. and M.I.K. acknowledge support from the ERC Advanced Grant 338957 FEMTO/NANO and from the NWO via the Spinoza Prize. GV acknowledges support from NSF Grant DMR-1406568.

\appendix

\section{The model}
We remind the reader that the model is defined by the following constitutive equations:
\begin{eqnarray} \label{SM_eq:continuity}
&& \partial_t \delta n_{{\bm K}} + n_0 {\bm \nabla}\cdot {\bm v}_{{\bm K}} = 0
~,
\nonumber\\
&& \partial_t \delta n_{{\bm K}'} + n_0 {\bm \nabla}\cdot {\bm v}_{{\bm K}'} = 0
~,
\end{eqnarray}
and
\begin{eqnarray} \label{SM_eq:Navier_Stokes}
&& \partial_t {\bm v}_{{\bm K}} + \omega_c {\hat {\bm z}} \times {\bm v}_{{\bm K}} + \frac{s^2}{n_0} {\bm \nabla}\delta n_{{\bm K}} - \frac{e}{m} {\bm \nabla} \phi = 0
~,
\nonumber\\
&& \partial_t {\bm v}_{{\bm K}'} - \omega_c {\hat {\bm z}} \times {\bm v}_{{\bm K}'} + \frac{s^2}{n_0} {\bm \nabla}\delta n_{{\bm K}'} - \frac{e}{m} {\bm \nabla} \phi = 0
~,
\end{eqnarray}
while the electrostatic potential satisfies
\begin{eqnarray} \label{SM_eq:potential}
\phi ({\bm r}) = e \int d^2{\bm r}' \frac{\delta n_{{\bm K}}({\bm r}') + \delta n_{{\bm K}'} ({\bm r}')}{|{\bm r}-{\bm r}'|}
~.
\end{eqnarray}
Since the electrons are confined in a half-plane, the system of Eqs.~(\ref{SM_eq:continuity})-(\ref{SM_eq:potential}) is a integro-differential system that can be solved by, e.g., the Wiener-Hopf technique. In what follows we assume translational invariance along the ${\hat {\bm y}}$ direction, and that all functions have a $\sim e^{-i(\omega t - q y)}$-dependence.

Fetter~\cite{Fetter_prb_1985} simplified the problem by introducing an approximation of Eq.~(\ref{SM_eq:potential}). He replaced it with the following differential equation
\begin{eqnarray} \label{SM_eq:potential_diff}
\partial_x^2 \phi(x) - 2 q^2 \phi(x) = 4\pi e |q| \big[ \delta n_{{\bm K}}(x) + \delta n_{{\bm K}'} (x) \big]
~.
\end{eqnarray}
The big advantage of such equation is that in this way one has to solve a system of ordinary (linear) differential equation.

Solving Eq.~(\ref{SM_eq:Navier_Stokes}) we find
\begin{eqnarray} \label{SM_eq:vel_K}
v_{{\bm K},x} &=& i \frac{s^2}{n_0} \frac{\omega \partial_x \delta n_{{\bm K}}(x) - q\omega_c \delta n_{{\bm K}}(x)}{\omega_c^2-\omega^2}
\nonumber\\
&-& i\frac{e}{m} \frac{\omega \partial_x\phi(x) - q \omega_c \phi(x)}{\omega_c^2-\omega^2}
~,
\nonumber\\
v_{{\bm K},y} &=& \frac{s^2}{n_0} \frac{\omega_c \partial_x \delta n_{{\bm K}}(x) - q\omega \delta n_{{\bm K}}(x)}{\omega_c^2-\omega^2}
\nonumber\\
&-& \frac{e}{m} \frac{\omega_c \partial_x\phi(x) - q \omega \phi(x)}{\omega_c^2-\omega^2}
~,
\end{eqnarray}
for the valley ${\bm K}$, and
\begin{eqnarray} \label{SM_eq:vel_K1}
v_{{\bm K}',x} &=& i \frac{s^2}{n_0} \frac{\omega \partial_x \delta n_{{\bm K}'}(x) + q\omega_c \delta n_{{\bm K}'}(x)}{\omega_c^2-\omega^2}
\nonumber\\
&-& i\frac{e}{m} \frac{\omega \partial_x\phi(x) + q \omega_c \phi(x)}{\omega_c^2-\omega^2}
~,
\nonumber\\
v_{{\bm K}',y} &=& -\frac{s^2}{n_0} \frac{\omega_c \partial_x \delta n_{{\bm K}'}(x) + q\omega \delta n_{{\bm K}'}(x)}{\omega_c^2-\omega^2}
\nonumber\\
&+& \frac{e}{m} \frac{\omega_c \partial_x\phi(x) + q \omega \phi(x)}{\omega_c^2-\omega^2}
~,
\end{eqnarray}
For the electrons in valley ${\bm K}'$. The main difference between Eq.~(\ref{SM_eq:vel_K}) and~(\ref{SM_eq:vel_K1}), besides the replacement $\delta n_{{\bm K}}(x) \to \delta n_{{\bm K}'}(x)$, is the sign of $\omega_c$. Plugging Eqs.~(\ref{SM_eq:vel_K})-(\ref{SM_eq:vel_K1}) into Eq.~(\ref{SM_eq:continuity}) we find
\begin{eqnarray} \label{SM_eq:continuity_2}
\big[s^2 (\partial_x^2 - q^2)- \omega_{\rm c}^2 + \omega^2 \big] \delta n_{\bm K}(x) = \frac{e n_0}{m} (\partial_x^2 - q^2) \phi(x)
~.
\end{eqnarray}
The same equation holds for $\delta n_{{\bm K}'}(x)$.
Note that no approximation has been done in Eqs.~(\ref{SM_eq:vel_K})-(\ref{SM_eq:continuity_2}), and they are therefore valid also in the description of the full model. A further simplification occurs by introducing the variables $\delta n_{{\bm K}}(x) = \big[n_{\rm sum}(x)+n_{\rm diff}(x)\big]/2$ and $\delta n_{{\bm K}'}(x) = \big[n_{\rm sum}(x)-n_{\rm diff}(x)\big]/2$. From Eq.~(\ref{SM_eq:continuity_2}) we get
\begin{equation} \label{SM_eq:continuity_p}
\big[s^2 (\partial_x^2 - q^2) +\omega^2 - \omega_c^2\big] n_{\rm sum}(x) = \frac{2 e n_0}{m} (\partial_x^2 - q^2) \phi(x)
~,
\end{equation}
and
\begin{eqnarray}\label{SM_eq:continuity_m}
\big[s^2 (\partial_x^2 - q^2) +\omega^2 - \omega_c^2\big] n_{\rm diff}(x) = 0 
~.
\end{eqnarray}
Note that Eq.~(\ref{SM_eq:continuity_p}) is to be solved together with Eq.~(\ref{SM_eq:potential}) [or its approximate counterpart~(\ref{SM_eq:potential_diff})]. On the contrary, $n_{\rm diff}(x)$ satisfies a differential equation which is independent of $\phi(x)$, and therefore it is exact whether or not the approximation~(\ref{SM_eq:potential_diff}) is considered. Its solution is given by
\begin{eqnarray} \label{SM_eq:evanescent_solution_n_diff}
n_{\rm diff}(x) = {\bar n}_{\rm diff} \exp(\kappa_- x)
~,
\end{eqnarray}
where
\begin{eqnarray}
\kappa_- = \sqrt{\frac{\omega_c^2 - \omega^2 + s^2 q^2}{s^2}}
~.
\end{eqnarray}

\section{The Wiener-Hopf solution}
\label{SM_app:WienerHopf}
We now derive the equation for the plasmon dispersion by the Wiener-Hopf technique. Taking the one-sided Fourier transform of Eqs.~(\ref{SM_eq:continuity})-(\ref{SM_eq:Navier_Stokes}), {\it i.e.} integrating them over $x$ between $-\infty$ and $0$ with the weight $e^{-i k x}$, we get
\begin{eqnarray} \label{SM_eq:cont_WH}
&&  \omega n_{\rm sum}(k) - 2 n_0 \big[ k v_{{\bm K},x}(k) + q v_{{\bm K},y}(k) \big] = -\frac{\omega {\bar n}_{\rm diff}}{-i k + \kappa_-}
~,
\nonumber\\
&&  \omega n_{\rm diff}(k) - 2 n_0 \big[ k v_{{\bm K}',x}(k) + q v_{{\bm K}',y}(k) \big] = \frac{\omega {\bar n}_{\rm diff}}{-i k + \kappa_-}
~,
\nonumber\\
\end{eqnarray}
and
\begin{widetext}
\begin{eqnarray} \label{SM_eq:NS_WH}
&& i \omega v_{{\bm K},x}(k) +\omega_c v_{{\bm K},y}(k) - \frac{i k s^2}{2 n_0} n_{\rm sum}(k) + \frac{i e k}{m} \phi(k) = \frac{s^2}{2n_0} {\bar n}_{\rm sum} + \frac{s^2 \kappa_-}{2 n_0(k_--i k)} {\bar n}_{\rm diff} - \frac{e}{m} {\bar \phi}
~,
\nonumber\\
&& i \omega v_{{\bm K},y}(k) -\omega_c v_{{\bm K},x}(k) - \frac{i q s^2}{2 n_0} n_{\rm sum}(k) + \frac{i e q}{m} \phi(k) = \frac{i q s^2}{2 n_0(k_--i k)} {\bar n}_{\rm diff}
~,
\nonumber\\
&& i \omega v_{{\bm K}',x}(k) -\omega_c v_{{\bm K}',y}(k) - \frac{i k s^2}{2 n_0} n_{\rm sum}(k) + \frac{i e k}{m} \phi(k) = \frac{s^2}{2n_0} {\bar n}_{\rm sum} - \frac{s^2 \kappa_-}{2 n_0(k_--i k)} {\bar n}_{\rm diff} - \frac{e}{m} {\bar \phi}
~,
\nonumber\\
&& i \omega v_{{\bm K}',y}(k) +\omega_c v_{{\bm K}',x}(k) - \frac{i q}{2 n_0} n_{\rm sum}(k) + \frac{i e q}{m} \phi(k) =  -\frac{i q s^2}{2 n_0(k_--i k)} {\bar n}_{\rm diff}
~.
\end{eqnarray}
in these equation we used that the velocity along the ${\hat {\bm x}}$ direction vanishes at the boundary $x=0$. Here ${\bar n}_{\rm sum}$ and ${\bar \phi}$ denote the value of the functions at $x=0$. Solving Eq.~(\ref{SM_eq:NS_WH}) and substituting into Eq.~(\ref{SM_eq:cont_WH}) we get the following equations:
\begin{eqnarray} \label{SM_eq:n_sum_n_diff_intermediate}
&& (k^2 + \kappa_-^2) n_{\rm sum}(k) - 2 \frac{e n_0}{m s^2} (k^2 + q^2) \phi(k) = i k \left[ {\bar n}_{\rm sum}  - 2 \frac{e n_0}{m s^2} {\bar \phi} \right]
 + q \frac{\omega_c}{\omega} {\bar n}_{\rm diff}
~,
\nonumber\\
&& {\bar n}_{\rm diff} = \frac{q}{\kappa_-}  \frac{\omega_c}{\omega} \left[{\bar n}_{\rm sum} - 2 \frac{e n_0}{m s^2} {\bar \phi} \right]
~,
\end{eqnarray}
from which we get
\begin{eqnarray} \label{SM_eq:n_plus_k_final}
n_{\rm sum}(k) =  2 \frac{e n_0}{m s^2} \frac{k^2 + q^2}{k^2 + \kappa_-^2} \phi(k) + \frac{1}{k^2 + \kappa_-^2} \left[ i k 
+\frac{q^2}{\kappa_-} \frac{\omega_c^2}{\omega^2} \right] \left[ {\bar n}_{\rm sum}  - 2 \frac{e n_0}{m s^2} {\bar \phi} \right]
~.
\end{eqnarray} 
We now extend $k$ to the whole complex plane, and we rename $\phi(k) \to \phi_+(k)$ and $n_{\rm sum}(k) \to n_+(k)$, where the ``$+$'' sign stands for the fact that the function is analytic in the upper-half of the complex plane. Recall indeed that Eq.~(\ref{SM_eq:n_sum_n_diff_intermediate}) was obtained by Fourier transforming over $x$ between $-\infty$ and $0$, and therefore the functions $n_{\rm sum}(k)$ and $\phi(k)$ that appear there are {\it by definition} analytic for $\Im m (k) >0$ .We complement Eq.~(\ref{SM_eq:n_plus_k_final}) with the Fourier transform of Eq.~(\ref{SM_eq:potential}). We get
\begin{eqnarray}
\left\{
\begin{array}{l}
{\displaystyle n_+(k) =  2 \frac{e n_0}{m s^2} \frac{k^2 + q^2}{k^2 + \kappa_-^2} \phi_+(k) + \frac{1}{k^2 + \kappa_-^2} \left[ i k 
+\frac{q^2}{\kappa_-} \frac{\omega_c^2}{\omega^2} \right] \left[ {\bar n}_{\rm sum}  - 2 \frac{e n_0}{m s^2} {\bar \phi} \right] }
\vspace{0.3cm}\\
{\displaystyle \phi_+(k)+\phi_-(k) = -\frac{2\pi e}{\sqrt{k^2+q^2}} n_+(k) }
\end{array}
\right.
~.
\end{eqnarray}
We therefore get the equation
\begin{eqnarray} \label{SM_eq:phi_equation}
\left(k^2 + \kappa_-^2 + \frac{4\pi e^2 n_0}{m s^2} \sqrt{k^2 + q^2} \right)\phi_+(k)+ (k^2 + \kappa_-^2)\phi_-(k) =  -\frac{2\pi e}{\sqrt{k^2+q^2}} \left[ i k 
+\frac{q^2}{\kappa_-}  \frac{\omega_c^2}{\omega^2} \right] \left[ {\bar n}_{\rm sum}  - 2 \frac{e n_0}{m s^2} {\bar \phi} \right]
~.
\nonumber\\
\end{eqnarray}
\end{widetext}
Eq.~(\ref{SM_eq:phi_equation}) can be solved with the Wiener-Hopf technique. 
The solution is shown in the main text.
We define
\begin{eqnarray}
G(k) &\equiv& 1 + 2 \alpha \frac{\sqrt{k^2 + q^2}}{k^2 + \kappa_-^2} 
\nonumber\\
&=&
\frac{G_+(k)}{G_-(k)}
~,
\end{eqnarray}
where $\alpha = 2 \pi e^2 n_0/(m s^2)$ and ($\eta \to 0^+$)~\cite{Noble_WH_book}
\begin{eqnarray}
G_{\pm}(k) = \exp \left[ \int_{-\infty\mp i\eta}^{\infty\mp i\eta} \frac{dz}{2\pi i} \frac{\ln G(z)}{z-k} \right]
~.
\end{eqnarray}
The function $G_+(k)$ [$G_-(k)$] is analytic in the upper (lower) half of the complex plane, including the real axis. Eq.~(\ref{SM_eq:phi_equation}) becomes
\begin{eqnarray} \label{SM_eq:phi_equation_2}
&&
G_+(k) \phi_+(k)+ G_-(k) \phi_-(k) =  
\nonumber\\
&&-\frac{2\pi e}{\sqrt{k^2+q^2}} \left[ i k 
+\frac{q^2}{\kappa_-} \frac{\omega_c^2}{\omega^2} \right] \frac{G_-(k)}{k^2 + \kappa_-^2} {\bar \Xi}
~.
\nonumber\\
\end{eqnarray}
Here ${\bar \Xi} = {\bar n}_{\rm sum}  - 2 e n_0 {\bar \phi}/(m s^2)$. The term on the second line of Eq.~(\ref{SM_eq:phi_equation_2}) is rewritten using the following equality:
\begin{eqnarray}
F(k) &\equiv& - \frac{1}{\sqrt{k^2+q^2}} \left[ i k 
+\frac{q^2}{\kappa_-} \frac{\omega_c^2}{\omega^2} \right] \frac{G_-(k)}{k^2 + \kappa_-^2}
\nonumber\\
&=&
F_+(k) + F_-(k)
~,
\end{eqnarray}
where $F_+(k)$ [$F_-(k)$] is analytic in the upper (lower) complex plane, including the real axis, and reads
\begin{eqnarray} \label{SM_eq:F_pm}
F_\pm (k) = \pm \int_{-\infty \mp i\eta}^{\infty \mp i \eta} \frac{dz}{2\pi i} \frac{F(z)}{z-k}
~.
\end{eqnarray}
Note that here $F(k)$ is defined differently with respect to the main text.
Eq.~(\ref{SM_eq:phi_equation_2}) can now be rewritten as
\begin{eqnarray}
&&
G_+(k) \phi_+(k) - 2\pi e F_+(k) {\bar \Xi} = 
\nonumber\\
&& 
2\pi eF_-(k) {\bar \Xi} - G_-(k) \phi_-(k)
~.
\nonumber\\
\end{eqnarray}
Since the left-hand side is analytic for $\Im m (k) \geq 0$ and the right-hand side is analytic for $\Im m(k)\leq 0$, together they define a function analytic in the whole complex plane. Since both sides have to vanish in the limit $|k|\to \infty$, we get
\begin{eqnarray}
\phi_+(k) = 2\pi e \frac{F_+(k)}{G_+(k)} {\bar \Xi}
~.
\end{eqnarray}
Plugging this equation back into Eq.~(\ref{SM_eq:n_plus_k_final}) we finally get
\begin{eqnarray} \label{SM_eq:n_plus_k_final_2}
n_+(k) =  \frac{\displaystyle 2 \alpha (k^2 + q^2) \frac{F_+(k)}{G_+(k)} +  i k 
+\frac{q^2}{\kappa_-} \frac{\omega_c^2}{\omega^2} }{k^2+\kappa_-^2} {\bar \Xi}
~.
\nonumber\\
\end{eqnarray} 
Since $n_+(k)$ must be analytic in $k = i \kappa_-$, the numerator on the right-hand side of Eq.~(\ref{SM_eq:n_plus_k_final_2}) must have a zero at that point. Therefore we get the equation
\begin{eqnarray} \label{SM_eq:WH_plasmon_equation}
2 \alpha (q^2 - \kappa_-^2) \frac{F_+(i\kappa_-)}{G_+(i\kappa_-)} - \kappa_- 
+\frac{q^2}{\kappa_-}  \frac{\omega_c^2}{\omega^2} = 0
~.
\end{eqnarray}

\section{The calculation of $G_\pm(k)$ and $F_+(k)$}
To determine the function $G_{\pm}(k)$ we first consider the function
\begin{eqnarray}
{\tilde G}(k) &\equiv& k^2 + \kappa_-^2 + 2 \alpha \sqrt{k^2 + q^2} 
\nonumber\\
&=&
\frac{{\tilde G}_+(k)}{{\tilde G}_-(k)}
~.
\end{eqnarray}
To determine the functions ${\tilde G}_{\pm}(k)$ we take the derivative of $\ln{\tilde G}(k)$ with respect to $k$. We get
\begin{eqnarray} \label{SM_eq:DlogG}
&& \partial_k \ln {\tilde G}(k) = \frac{\partial_k {\tilde G}_+(k)}{{\tilde G}_+(k)} - \frac{\partial_k {\tilde G}_-(k)}{{\tilde G}_-(k)} 
\nonumber\\
&& 
=
\frac{2 k}{ k^2 + \kappa_-^2 + 2 \alpha \sqrt{k^2 + q^2}} \left( 1 + \frac{\alpha}{\sqrt{k^2+q^2}} \right)
~.
\nonumber\\
\end{eqnarray}
Using Eq.~(\ref{SM_eq:F_pm}) we can now split the second line of Eq.~(\ref{SM_eq:DlogG}) in the sum of two functions, one analytic in the upper half of the complex plane and the other analytic in its lower half. We identify them with $\partial_k {\tilde G}_+(k)/{\tilde G}_+(k)$ and $-\partial_k {\tilde G}_-(k)/{\tilde G}_-(k)$, respectively. Furthermore, since the function $\partial_k \ln {\tilde G}(k)$ is odd in $k$,
\begin{eqnarray}
\frac{\partial_k {\tilde G}_-(k)}{{\tilde G}_-(k)} &=&
\int_{-\infty + i\eta}^{\infty + i \eta} \frac{dz}{2\pi i} \frac{\partial_z \ln {\tilde G}(z)}{z-k} 
\nonumber\\
&=&
-\int_{-\infty - i\eta}^{\infty - i \eta} \frac{dz}{2\pi i} \frac{\partial_z \ln {\tilde G}(-z)}{z+k} 
\nonumber\\
&=&
\left. \frac{\partial_k {\tilde G}_+(k)}{{\tilde G}_+(k)} \right|_{k\to -k}
~,
\end{eqnarray}
and therefore we need to compute only one of the two functions. We get
\begin{eqnarray} \label{SM_eq:G_p_tilde_intermezzo}
\frac{\partial_k {\tilde G}_+(k)}{{\tilde G}_+(k)} &=&
\int_{-\infty}^{\infty} \frac{dz}{2\pi i} \frac{1}{z-(k+i\eta)} 
\left( 1 + \frac{\alpha}{\sqrt{z^2+q^2}} \right)
\nonumber\\
&\times&
\frac{2 z}{ z^2 + \kappa_-^2 + 2 \alpha \sqrt{z^2 + q^2}} 
~.
\end{eqnarray}
To avoid the pole $z= k+i\eta$, we close the contour in the lower half of the complex plane, circulating in the clockwise direction. The contour excludes the branch cut of the square root, which runs from $z=-iq$ to $z=-i \infty$. Note that the denominator of the last line on Eq.~(\ref{SM_eq:G_p_tilde_intermezzo}) has no poles thanks to properties of the square root, namely that it changes sign on the two sides of the branch cut.
After some straightforward manipulations we get
\begin{eqnarray}
\frac{\partial_k {\tilde G}_+(k)}{{\tilde G}_+(k)} &=&
\frac{2 \alpha}{\pi} \int_q^\infty d x \frac{x}{k+ix} \frac{1}{\sqrt{x^2-q^2}}
\nonumber\\
&\times&
\frac{x^2 + \kappa_-^2 - 2 q^2}{(x^2-\kappa_-^2)^2 + 4\alpha^2(x^2-q^2)}
~,
\end{eqnarray}
and therefore, after an integration over $k$,
\begin{eqnarray} \label{SM_eq:lnGtildepm_final}
\ln {\tilde G}_+(k) &=&
\frac{2 \alpha}{\pi} \int_q^\infty d x 
\frac{x^2 + \kappa_-^2 - 2 q^2}{(x^2-\kappa_-^2)^2 + 4\alpha^2(x^2-q^2)} 
\nonumber\\
&\times&
\frac{x}{\sqrt{x^2-q^2}} \ln(k+ix)
~,
\nonumber\\
\ln {\tilde G}_-(k) &=&
-\frac{2 \alpha}{\pi} \int_q^\infty d x 
\frac{x^2 + \kappa_-^2 - 2 q^2}{(x^2-\kappa_-^2)^2 + 4\alpha^2(x^2-q^2)} 
\nonumber\\
&\times&
\frac{x}{\sqrt{x^2-q^2}} \ln(k-ix)
~.
\end{eqnarray}
The function $G_+(k)$ [$G_-(k)$] can be found from Eq.~(\ref{SM_eq:lnGtildepm_final}) by subtracting [adding] $\ln(k+i\kappa_-)$ [$\ln(k-i\kappa_-)$]. Using the fact that (in the limit $q\ll \kappa_-$)
\begin{eqnarray}
\frac{2 \alpha}{\pi} \int_q^\infty d x 
\frac{x^2 + \kappa_-^2 - 2 q^2}{(x^2-\kappa_-^2)^2 + 4\alpha^2(x^2-q^2)} 
\frac{x}{\sqrt{x^2-q^2}} 
=1
~,
\nonumber\\
\end{eqnarray}
we finally get
\begin{eqnarray} \label{SM_eq:Gpm_final}
G_+(k) &=&
\exp \Bigg[ \frac{2 \alpha}{\pi} \int_q^\infty d x 
\frac{x^2 + \kappa_-^2 - 2 q^2}{(x^2-\kappa_-^2)^2 + 4\alpha^2(x^2-q^2)} 
\nonumber\\
&\times&
\frac{x}{\sqrt{x^2-q^2}} \ln\left(\frac{k+ix}{k+i\kappa_-}\right)
\Bigg]
~,
\nonumber\\
G_-(k) &=& \exp \Bigg[
-\frac{2 \alpha}{\pi} \int_q^\infty d x 
\frac{x^2 + \kappa_-^2 - 2 q^2}{(x^2-\kappa_-^2)^2 + 4\alpha^2(x^2-q^2)} 
\nonumber\\
&\times&
\frac{x}{\sqrt{x^2-q^2}}  \ln\left(\frac{k-ix}{k-i\kappa_-}\right)
\Bigg]
~.
\end{eqnarray}
Note that $G_\pm(k) \to 1$ in the limit $k\to \infty$.

We are now in the position to calculate the function $F_+(k)$, which reads
\begin{eqnarray} \label{SM_eq:F_k_contour}
F_+(k) &=&
-\int_{-\infty}^{\infty} \frac{dz}{2\pi i} \frac{(z^2+q^2)^{-1/2}}{z-(k+i\eta)}
\left[ i z  +\frac{q^2}{\kappa_-} \frac{\omega_c^2}{\omega^2} \right] 
\nonumber\\
&\times&
\frac{G_-(z)}{z^2 + \kappa_-^2}
~.
\nonumber\\
\end{eqnarray}
Once again the contour is closed in the lower half of the complex plane and goes around the branch cut of the square root. Note that for small $q$ the pole $z = -i\kappa_-$ lies inside such branch cut, and that the function $G_-(z)$. is analytic on it. After some straightforward algebra we get
\begin{eqnarray}
F_+(k) &=&
\frac{i}{\pi} {\cal P} \int_q^\infty \frac{d y}{iy + k}\left(\frac{q^2}{\kappa_-} \frac{\omega_c^2}{\omega^2} + y\right) \frac{1}{y^2 - \kappa_-^2}
\nonumber\\
&\times&
\frac{G_-(-i y)}{\sqrt{y^2 - q^2}} 
~,
\end{eqnarray}
where ${\cal P}$ stands for the principal value. Finally, the functions that enter Eq.~(\ref{SM_eq:WH_plasmon_equation}) are
\begin{eqnarray}
G_+(i\kappa_-) &=&
\exp \Bigg[ \frac{2 \alpha}{\pi} \int_q^\infty d x 
\frac{x^2 + \kappa_-^2 - 2 q^2}{(x^2-\kappa_-^2)^2 + 4\alpha^2(x^2-q^2)} 
\nonumber\\
&\times&
\frac{x}{\sqrt{x^2-q^2}} \ln\left(\frac{\kappa_-+x}{2\kappa_-}\right)
\Bigg]
~,
\nonumber\\
F_+(i\kappa_-) &=&
\frac{1}{\pi} {\cal P} \int_q^\infty \frac{d y}{y + \kappa_-}\left(\frac{q^2}{\kappa_-} \frac{\omega_c^2}{\omega^2} + y\right) \frac{(y^2 - q^2)^{-1/2}}{y^2 - \kappa_-^2}
\nonumber\\
&\times&
\exp \Bigg[
-\frac{2 \alpha}{\pi} \int_q^\infty d x 
\frac{x^2 + \kappa_-^2 - 2 q^2}{(x^2-\kappa_-^2)^2 + 4\alpha^2(x^2-q^2)} 
\nonumber\\
&\times&
\frac{x}{\sqrt{x^2-q^2}}  \ln\left(\frac{y+x}{y+\kappa_-}\right)
\Bigg]
~.
\end{eqnarray}
In the limit $q\to 0$, and assuming that $\omega = c q$, Eq.~(\ref{SM_eq:WH_plasmon_equation}) becomes
\begin{eqnarray} \label{SM_eq:WH_plasmon_equation_app}
\frac{s^2}{c^2}  - 1 = 2 {\bar \alpha} \frac{f}{g}
~,
\end{eqnarray}
where we defined ${\bar \alpha} \equiv s \alpha/\omega_c = 2 \pi e^2 n_0/(m s\omega_c)$, and
\begin{eqnarray}
g &=&
\exp \Bigg[ \frac{2 {\bar \alpha}}{\pi} \int_0^\infty d x 
\frac{x^2 + 1}{(x^2-1)^2 + 4{\bar \alpha}^2 x^2} 
\ln\left(\frac{1+x}{2}\right)
\Bigg]
~,
\nonumber\\
f &=&
\frac{1}{\pi} {\cal P} \int_{{\bar q}_c}^\infty \frac{d y}{y + 1}\left(\frac{s^2}{c^2} + y\right) \frac{y^{-1}}{y^2 - 1}
\nonumber\\
&\times&
\exp \Bigg[
-\frac{2 {\bar \alpha}}{\pi} \int_0^\infty d x 
\frac{ \displaystyle (x^2 + 1) \ln\left(\frac{y+x}{y+1}\right) }{(x^2-1)^2 + 4{\bar \alpha}^2 x^2} 
\Bigg]
~.
\nonumber\\
\end{eqnarray}
The integral on the second line is infrared divergent and it is cut-off by the dimensionless screening length ${\bar q}_c= N_{\rm f} \alpha_{\rm ee} s k_{\rm F}/\omega_c$.

\section{The ratio ${\bar n}_{\rm sum}/{\bar n}_{\rm diff}$}
We first observe that, according to the second line of Eq.~(\ref{SM_eq:n_sum_n_diff_intermediate}), ${\bar n}_{\rm diff} = {\bar \Xi} c^{-1}$, in the limit $q\to 0$. To calculate ${\bar n}_{\rm sum}$ we have to consider the Fourier transform of  $n_+(k)$, given by Eq.~(\ref{SM_eq:n_plus_k_final_2}), in the limit $x\to 0^-$. This implies that the contour of integration is to be closed in the lower half of the complex plane, where the functions $F_+(k)$ and $G_+(k)$ have branch cuts. Prior to performing the Fourier transform of $n_+(k)$, it is convenient to rewrite $F_+(k) = F(k) + F^{(1)}(k)$, where (hereafter, to simplify the results, we neglect contributions that would vanish in the limit $q\to 0$)
\begin{eqnarray}
F^{(1)}(k) &=& \frac{1}{\pi} \int_q^\infty dx \frac{(x^2-q^2)^{-1/2}}{x+ik} \left[ \frac{q^2}{\kappa_-} \frac{\omega_c^2}{\omega^2} - x \right] 
\nonumber\\
&\times&
\frac{\Re e \big[G_-(ix + 0^+)\big]}{k_-^2-x^2}
~.
\end{eqnarray}
To obtain this expression we closed the contour of integration of Eq.~(\ref{SM_eq:F_k_contour}) in the upper half of the complex plane. Note that $F^{(1)}(k)$ is analytic for $k \to -i k$. We also note that
\begin{eqnarray}
&& 
G_+(-ik+0^+) = (\kappa_- - k)^{-1}
\exp \Bigg\{ \frac{2 \alpha}{\pi} \int_q^\infty d x 
\frac{x}{\sqrt{x^2-q^2}} 
\nonumber\\
&& 
\times
\frac{x^2 + \kappa_-^2 - 2 q^2}{(x^2-\kappa_-^2)^2 + 4\alpha^2(x^2-q^2)} 
\big[ \ln|x-k| - i \pi \Theta(k-x)\big]
\Bigg\}
~,
\nonumber\\
\end{eqnarray}
and
\begin{eqnarray}
&&
G_-(ik+0^+) = (k - \kappa_-) \exp \Bigg\{
-\frac{2 \alpha}{\pi} \int_q^\infty d x 
\frac{x}{\sqrt{x^2-q^2}}  
\nonumber\\
&&
\times
\frac{x^2 + \kappa_-^2 - 2 q^2}{(x^2-\kappa_-^2)^2 + 4\alpha^2(x^2-q^2)} 
\big[\ln|k-x| - i \pi \Theta(x-k)\big]
\Bigg\}
~.
\nonumber\\
\end{eqnarray}

\section{The solution of the approximate model}

The set of differential equations given by Eqs.~(\ref{SM_eq:potential_diff}),~(\ref{SM_eq:continuity_p}) and~(\ref{SM_eq:continuity_m}) can be easily solved. We look for a solution of the form $n_{\rm sum}(x) = {\bar n}_{\rm sum} e^{\kappa_+ x}$ and  $\phi(x) = {\bar \phi} e^{\kappa_+ x}$, and we set to zero the determinant of the resulting linear system. Therefore
\begin{equation} \label{SM_eq:kappa_eq_1}
4 \Omega_q^2 (\kappa_+^2 - q^2) - (\kappa_+^2 - 2 q^2) \big[s^2(\kappa_+^2-q^2) +\omega^2 - \omega_c^2\big]=0
~,
\end{equation}
where $\Omega_q^2 = 2\pi e^2 n_0 q/m$ is the bulk plasmon frequency at zero magnetic field.
Eq.~(\ref{SM_eq:kappa_eq_1}) has two solutions, which we denote by $\kappa_{+,1}$ and $\kappa_{+,2}$. We write the solution of the system of differential equations as
\begin{eqnarray} \label{SM_eq:solution_evanescent}
\phi(x) = {\bar \phi}_1 e^{\kappa_{+,1} x} + {\bar \phi}_2 e^{\kappa_{+,2} x}
~,
\end{eqnarray}
and
\begin{eqnarray} \label{SM_eq:solution_evanescent_dens}
n_{\rm sum}(x) &=& \frac{(\kappa_{+,1}^2 - 2q^2) {\bar \phi}_1}{4\pi e q} e^{\kappa_{+,1} x} 
\nonumber\\
&+&
\frac{(\kappa_{+,2}^2 - 2q^2) {\bar \phi}_2}{4\pi e q} e^{\kappa_{+,2} x}
~,
\end{eqnarray}
and we impose the boundary conditions to find an equation for $\kappa_{+,1}$ and $\kappa_{+,2}$, which, together with Eq.~(\ref{SM_eq:kappa_eq_1}), allows us to determine the edge plasmon frequencies. Such boundary conditions are given by (i) the continuity of the potential and (ii) of its derivative, and the vanishing of (iii) $v_{{\bm K},x}$ and (iv) $v_{{\bm K}',x}$ at the boundary $x=0$. The potential for $x>0$ is given by $\phi(x) = {\bar \phi}_0 e^{- \sqrt{2} q x}$, while all other functions of course vanish. The boundary conditions form a system of four linear homogeneous equations in the four unknowns ${\bar \phi}_0$, ${\bar \phi}_1$, ${\bar \phi}_2$, and ${\bar n}_{\rm diff}$. This system has a non-trivial solution if its determinant is zero, {\it i.e.} if
\begin{eqnarray} \label{SM_eq:kappa_eq_2}
&& \omega^2 s^2 \kappa_- \kappa_{+,1} \kappa_{+,2} (\kappa_{+,1} + \kappa_{+,2}) 
\nonumber\\
&&
-\sqrt{2} q \kappa_- \omega^2 \big[s^2 (\kappa_{+,1}^2 + \kappa_{+,1} \kappa_{+,2} + \kappa_{+,2}^2) - 4 \Omega_q^2\big] 
\nonumber\\
&&
- 2 s^2 q^4 \omega_c^2 + q^2 \omega_c^2 (s^2 \kappa_{+,1} \kappa_{+,2} +  4 \Omega_q^2) 
\nonumber\\
&&
+ \sqrt{2} s^2 q^3 (2 \kappa_- \omega^2 + (\kappa_{+,1} +\kappa_{+,2}) \omega_c^2) = 0
~.
\end{eqnarray}
Once Eq.~(\ref{SM_eq:kappa_eq_1}) has been solved and the resulting $\kappa_{+,1}$ and $\kappa_{+,2}$, together with $\kappa_-$, have been plugged into Eq.~(\ref{SM_eq:kappa_eq_2}), we can look for a low-energy solution of the form $\omega = c q$. Expanding Eq.~(\ref{SM_eq:kappa_eq_2}) for small $q$, and setting the coefficient of the lowest term of the Taylor expansion to zero we find
\begin{eqnarray} \label{SM_eq:sound_vel_approx}
c = s \sqrt{1 + 2\sqrt{2} \frac{2 \pi e^2 n_0}{m s \omega_c}}
~.
\end{eqnarray}
it is now interesting to study the density distribution for this mode. Plugging $\omega = c q$, with $c$ given by Eq.~(\ref{SM_eq:sound_vel_approx}), into Eqs.~(\ref{SM_eq:evanescent_solution_n_diff}),~(\ref{SM_eq:solution_evanescent}) and~(\ref{SM_eq:solution_evanescent_dens}) we find that, in the limit $\omega_c\to \infty$ and $q\to 0$,  ${\bar \phi}_1 = 0$, $\kappa_{2,+} = \kappa_- = \omega_c/s$, and
\begin{eqnarray}
&& n_{\rm sum}(x) = {\bar n}_{\rm diff} e^{\omega_c x/s}
~,
\nonumber\\
&& n_{\rm diff}(x) = {\bar n}_{\rm diff} e^{\omega_c x/s}
~,
\end{eqnarray}
and therefore $\delta n_{{\bm K}'}(x) = 0$ and $\delta n_{{\bm K}}(x) \neq 0$. Therefore, the edge plasmon contains only the contribution of the electron in valley ${\bm K}$. Reversing the sign of $\omega_c$ (or considering the plasmon with negative group velocity) we would get an edge plasmon whose contribution comes mainly from valley ${\bm K}'$.

\end{document}